%% file: main.tex
\documentclass[conference]{IEEEtran}
\usepackage[utf8]{inputenc}
\usepackage{times,graphics, dsfont, epsfig,amsmath,xspace,endnotes,pifont,multirow,rotating,listings,amssymb,algorithmic,color,caption,nicefrac,adjustbox,tabularx,mathtools, algorithmic,algorithm}
\usepackage{graphicx}
\usepackage{tabularx}

\newcolumntype{L}[1]{>{\raggedright\arraybackslash}p{#1}} 
\newcolumntype{C}[1]{>{\centering\arraybackslash}p{#1}} 
\newcolumntype{R}[1]{>{\raggedleft\arraybackslash}p{#1}} %
\usepackage{steinmetz}
\usepackage{color}
\usepackage[dvipsnames]{xcolor}
\usepackage{amsmath}
\usepackage{amssymb}
\usepackage{subcaption}
\usepackage{graphicx}
\usepackage[letterpaper, left=0.62in, right=0.62in, bottom=1in, top=0.75in]{geometry}
\usepackage{mathtools}
\DeclarePairedDelimiter{\ceil}{\lceil}{\rceil}

\usepackage{tikz}
\usepackage{amsthm}
\newtheorem{thm}{Theorem}

\newtheorem{defn}[thm]{Definition}

\renewcommand{\H} {{\bf{H}}}

\newcommand{\B} {{\bf{B}}}
\newcommand{\F} {{\bf{F}}}

\renewcommand{\b} {{\bf{b}}}

\renewcommand{\t} {{\bf{t}}}

\newcommand{\f} {{\bf{f}}}

\newcommand{\h} {{\bf{h}}}

\renewcommand{\r} {{\bf{r}}}

\newcommand{\s} {{\bf{s}}}
\newcommand{\y} {{\bf{y}}}
\newcommand{\w} {{\bf{w}}}
\newcommand{\Y} {{\bf{Y}}}
\newcommand{\x} {{\bf{x}}}
\newcommand{\X} {{\bf{X}}}
\newcommand{\Z} {{\bf{Z}}}
\newcommand{\R} {{\bf{R}}}
\newcommand{\z} {{\bf{z}}}
\newcommand{\T} {{\bf{T}}}

\newcommand{\W} {{\bf{W}}}
\newcommand{\Q} {{\bf{Q}}}

\newcommand{\nariman}[1]{\textcolor{red}{#1}}
\newcommand{\amir}[1]{\textcolor{blue}{#1}}

\author{
\IEEEauthorblockN{Nariman Torkzaban}
\IEEEauthorblockA{\textit{University of Maryland, College Park}\\
College Park, MD\\
narimant@umd.edu}
\and
\IEEEauthorblockN{Mohammad A. (Amir) Khojastepour}
\IEEEauthorblockA{\textit{NEC Laboratories,
America}\\
Princeton, NJ\\
amir@nec-labs.com}
\and
\IEEEauthorblockN{John S. Baras}
\IEEEauthorblockA{\textit{University of Maryland, College Park}\\
College Park, MD\\
baras@umd.edu}
}

\def\BibTeX{{\rm B\kern-.05em{\sc i\kern-.025em b}\kern-.08em
    T\kern-.1667em\lower.7ex\hbox{E}\kern-.125emX}}

\begin{document}

%\title{Channel Reciprocity Calibration for Co-operative Hybrid Beamforming in Distributed MIMO Systems}
% \title{Channel Reciprocity Calibration for Co-operative Hybrid Beamforming in Distributed MIMO Systems}

\title{Enabling Cooperative Hybrid Beamforming in TDD-based Distributed MIMO Systems}

% \title{Joint Synchronization \& Channel Reciprocity Calibration in Co-operative Multi-point TDD Systems}
% \title{On the Trade-off between Scanning Coverage and Uncertainty Region in  Beam Alignment for Multi-path Environments}
% Multi-user Beam Alignment for Millimeter Wave Systems in Multi-path Environments(ASILOMAR)
%\title{RIS-enabled Coverage by Dual-beam design in mmWave communication}
%\title{Shaping mmWave Wireless Channel via Composite Beamforming Design}
%\title{RIS Discussion: Dual-beam design for covering blind spots in mmWave communication using RIS }
%\title{Virtualized Network Function Placement\\for Cellular Network Slicing}



\maketitle
% \copyrightnotice
\begin{abstract}
Distributed massive MIMO networks are envisioned to realize cooperative multi-point transmission in next-generation wireless systems. For efficient cooperative hybrid beamforming, the cluster of access points (APs) needs to obtain precise estimates of the uplink channel to perform reliable downlink precoding. However, due to the radio frequency (RF) impairments between the transceivers at the two en-points of the wireless channel, full channel reciprocity does not hold which results in performance degradation in the cooperative hybrid beamforming (CHBF) unless a suitable reciprocity calibration mechanism is in place. We propose a two-step approach to calibrate any two hybrid nodes in the distributed MIMO system. We then present and utilize the novel concept of \emph{reciprocal tandem} to propose a low-complexity approach for jointly calibrating the cluster of APs and estimating the downlink channel. Finally, we validate our calibration technique's effectiveness through numerical simulation.
\end{abstract}

\begin{IEEEkeywords}
Channel Reciprocity Calibration, Distributed MIMO System, Collaborative Hybrid Beamforming.
\end{IEEEkeywords}

% \fi
\IEEEpeerreviewmaketitle

\input{introduction1.tex}

\input{problem_description.tex}

\input{reciprocal}
\input{downlink}

\input{evaluation.tex}

\input{conclusion.tex}

% \input{appendix}
% \input{L_free_formulation.tex}


\renewcommand{\nariman}[1]{\textcolor{red}{#1}}
\renewcommand{\amir}[1]{\textcolor{blue}{#1}}



%\bibliographystyle{IEEEtran}

\bibliographystyle{IEEEtran}
\bibliography{bibliography}

\end{document}

%% file: introduction1.tex
\section{Introduction}

Massive multiple-input multiple-output (MIMO) is to be widely used in the physical layer of the next-generation wireless systems, due to its potential for increasing the network capacity and user bit rates \cite{Wei22}. For this to be practical on the large scale, effective techniques are required to reduce the hardware cost and power consumption of such systems. A hybrid beamforming transceiver is an effective solution for low-cost massive MIMO by reducing the number of radio-frequency(RF) chains and hence expensive elements such as digital-to-analog/analog-to-digital converters (DAC/ADC), mixers, etc., and introducing simple phase shifters \cite{Nie22}. To mitigate the inter-cell interference resulting from the densification of the networks deploying large-scale massive MIMO, the distributed massive MIMO (a.k.a. cell-free massive MIMO) technology is proposed. 

A distributed massive MIMO system is an implementation of the cooperative multi-point transmission and is envisioned as a candidate for realizing 6G multi-antenna systems. It is comprised of a cluster of geographically distributed multi-antenna access points (APs) that collaborate toward serving a group of mobile users (MUs). To enable hybrid beamforming to an MU, the APs involved in the downlink transmission require knowledge of the downlink channel state information (CSI) to perform reliable downlink precoding. If the downlink and uplink are at the same frequency, exploiting the reciprocity of the uplink-downlink channel, the uplink estimates can be utilized for estimating the downlink CSI. Therefore, the recommended scheme for distributed MIMO systems is a time-division duplex(TDD). In practice with hybrid transceivers, the CSI not only depends on the wireless channel but also is a function of the RF response of the digital (DAC in the transmit and ADC in the receive direction) and the analog (power amplifier(PA) in the transmit and low-noise amplifier (LNA) in the receive direction) chains and the phase shift network (PSN) in between them. However, due to the asymmetry between the responses of the DAC/ADC and LNA/PA, full channel reciprocity does not hold. Therefore, channel reciprocity calibration is required to enable accurate downlink CSI acquisition. Various calibration designs such as bidirectional signaling between the APs and the MUs \cite{Kalten10}, and internal calibration at the APs \cite{She12}\cite{Vie14} are proposed in the literature for the full-digital transceivers. But none of these methods are practical for calibrating a hybrid transceiver, given the structure of the digital and the analog chain. Reciprocity calibration for hybrid beamforming in TDD-based massive MIMO systems was first addressed in \cite{Jiang18} where the authors divide the antenna array into two sub-arrays, and then perform internal calibration. However, their method is only practical for a subarray-based PSN. As opposed to this work, in this paper, we consider a fully-connected PSN. Calibration of a fully-connected PSN is studied in \cite{Wei20}. Within the context of distributed massive MIMO systems, the authors in \cite{Vie21} proposed a maximum likelihood (ML)-based calibration relying on joint beam sweep by all APs in the network. The authors consider a single mismatch parameter per AP and only consider fully-digital beamforming. However, in this paper,  similar to \cite{tork23} we consider a single parameter per chain in the hybrid structure. The main contributions of the paper are as follows:
\begin{itemize}
    \item We present a novel two-step low-complexity scheme for calibrating any two nodes in a distributed MIMO system. 
    \item We introduce the novel concept of \emph{reciprocal tandem} that is a cornerstone in the joint reciprocity calibration of the APs in a distributed MIMO system.
    \item Utilizing the concept of reciprocal tandem, we augment the two-step approach for reciprocity calibration of two nodes with a third step, and propose a low-complexity three-step scheme for the joint calibration of the cluster of APs in the distributed MIMO network.
    % \item We demonstrate how our reciprocity calibration technique enables the cluster of APs to perform the co-operative hybrid beamforming task by means of numerical simulation. 
\end{itemize}

The remainder of the paper is organized as follows. Section~\ref{sec:desc} describes the system model. In Section~\ref{sec:reciprocal tandem} we introduce the novel notion of reciprocal tandem. We then elaborate on the use of reciprocal tandem in performing reciprocity calibration between multiple APs and the corresponding downlink channel estimation for cooperative communications from distributed APs in Section~\ref{channel estimation}. Section~\ref{sec:evaluation} presents our evaluation results, and finally, in Section~\ref{sec:conclusions}, we highlight our conclusions.

%% file: problem_description.tex
\section{System model} 
\label{sec:desc}

Consider a distributed mmWave MIMO system where a cluster of $K$ co-operative multi-antenna APs employing hybrid beamforming, jointly serve a group of mobile users (MUs).

\subsection{Hybrid Beamforming Model}
\begin{figure}[!t]
\centering
\includegraphics[width=0.42\textwidth]{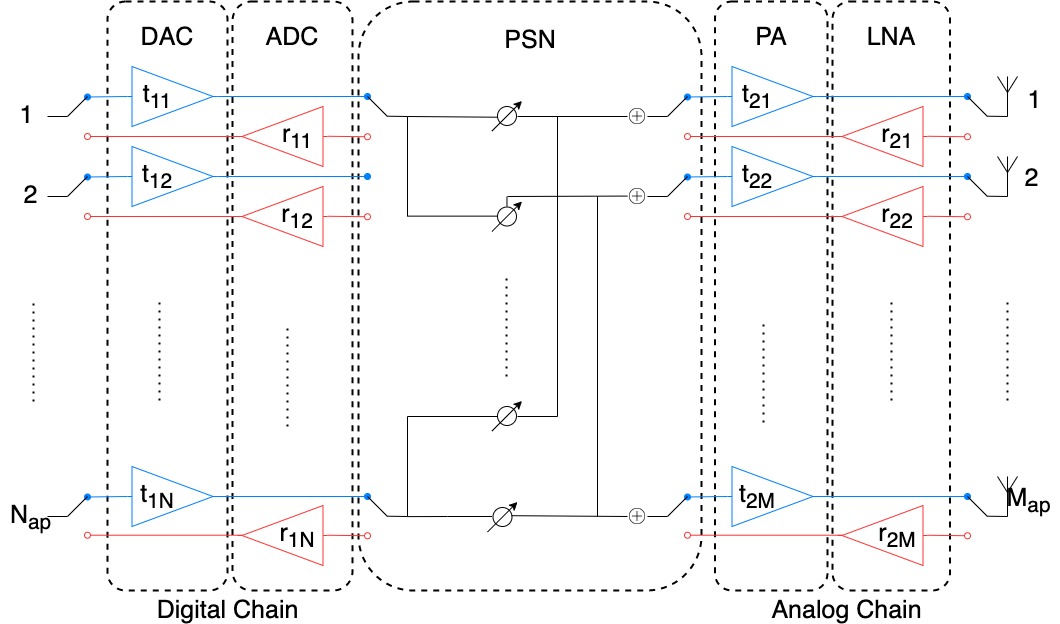}
  \caption{Hybrid Beamforming Model}
    \label{fig:hybmodel}
\end{figure}

The structure of a hybrid beamformer is shown in fig.~\ref{fig:hybmodel}. The hybrid beamformer at each AP is comprised of two stages that perform \emph{digital} and \emph{analog} beamforming, respectively. For each transmit stream, the output of the first-stage digital beamformer passes through the $N_{\text{ap}}$ digital RF chains. Then, the output of the digital RF chains is fed to the second-stage analog beamformer, followed by $M_{\text{ap}}$ analog RF chains each of which is connected to an antenna element where the antenna elements are arranged in a uniform linear array (ULA) structure. We assume the analog beamformer is realized through a fully-connected phase-shift network (PSN) comprised of phased-only vectors. This path is traversed in the reverse direction for a received stream. 
At the MUs, there are $M_{\text{mu}}$ and $N_{\text{mu}}$ analog and digital RF chains, respectively.

\subsection{Channel Model}
We assume a multipath mmWave channel $\H \in \mathbb{C}^{M_{\text{ap}}\times M_{\text{mu}}}$ between the MU and the AP with only a few spatial clusters with $L \ll N_{\text{ap}}, N_{\text{mu}}$ being the number of multipath components. The effective uplink (UL) and the downlink (DL) channels are given by, 
\begin{align}
    & \H_{DL} = \R_{\text{mu},1}\B^T\R_{ \text{mu},2}\H\T_{ \text{ap},2}\F\T_{\text{ap},1}, \\ & \H_{UL} = \R_{\text{ap},1}\F^T\R_{\text{ap},2}\H^T\T_{\text{mu},2}\B\T_{\text{mu},1}
\end{align}
where for $n\in\{\text{ap}, \text{mu}\}$, $\T_{ \text{n},1}, \R_{ \text{n},1} \in \mathbb{C}^{N_{\text{n}} \times N_{\text{n}}}$ represent the \emph{digital reciprocity calibration matrices} denoting the frequency responses of the transmit and receive digital RF chains (DAC/ADC) at node $n$, $\T_{\text{n},2}, \R_{\text{n},2} \in \mathbb{C}^{M_{\text{n}} \times M_{\text{n}}}$ are the \emph{analog reciprocity calibration matrices} denoting the transmit and receive frequency responses of the analog chains (PA/LNA) at node $n$, respectively. We note that all these matrices are diagonal with the diagonal entries modeling the gain and phase characteristics of each of the chain elements. The off-diagonal entries model the cross-talk between the RF hardware that is assumed to be almost zero under proper RF design \cite{Jiang15}. We assume that all the transceivers store codebooks for the hybrid beamforming task where each codebook consists of codewords that each represent a beamforming vector. The $\F \in \mathbb{C}^{M_{\text{ap}}\times N_{\text{ap}}}$ and $\B \in \mathbb{C}^{M_{\text{mu}}\times N_{\text{mu}}}$ are known as the beamforming matrices at the AP and the MU, respectively where each beamforming matrix consists of $N_t$ beamforming vectors. The matrices $\B$ and $\F$ model the beamformers that precode the input analog streams that are then amplified and sent to the analog chains. We note that each AP or MU uses the same matrices for beamforming both when transmitting and receiving. The transmission in the DL/UL direction can be modeled by, 
\begin{align}
    & \y_{\text{DL}} = \H_{\text{DL}}\x_{\text{DL}} + \z_{\text{DL}} \label{DL_tran}\\
    & \y_{\text{UL}} = \H_{\text{UL}}\x_{\text{UL}} + \z_{\text{UL}} \label{UL_tran}
\end{align}
where $\x_{{\text{DL}}} \in \mathbb{C}^{N_\text{ap}}$ and $\x_{{\text{UL}}} \in \mathbb{C}^{N_\text{mu}}$ are the input streams to the digital RF chains (i.e., the output of the baseband processing unit (BBU) which could, in turn, be digitally precoded symbols) at the AP and the MU, respectively. Similarly, $\z_{\text{DL}} \in \mathbb{C}^{M_{\text{mu}}}$ and $\z_{\text{UL}} \in \mathbb{C}^{M_{\text{ap}}}$ are the vectors of the additive white Gaussian noise (AWGN) with the distributions $\mathbf{z_{\text{DL}}} \sim \mathcal{C} \mathcal{N}\left(\mathbf{0}, \sigma_{\mathrm{z}}^{2} \mathbf{I}_{M_{\text{mu}}}\right)$ and $\mathbf{z_{\text{UL}}} \sim \mathcal{C} \mathcal{N}\left(\mathbf{0}, \sigma_{\mathrm{z}}^{2} \mathbf{I}_{M_{\text{ap}}}\right)$.

% In the DL direction, the AP may use a digital precoder $\W \in \mathbb{C}^{N_{\text{ap}} \times N_s}$, such that $\x_{\text{DL}} = \W\s$ where $\s \in \mathbb{C}^{N_s}$ is the vector of digital symbols transmitted from the AP with $N_s \leq N_{\text{ap}}$ being the number of data streams, $\y_{\text{DL}} \in \mathbb{C}^{M_{\text{mu}}}$ and $\y_{\text{UL}} \in \mathbb{C}^{M_{\text{ap}}}$ are the vectors of the received signal streams at the BBU in the DL and UL directions, respectively. 

%% file: reciprocal.tex
\section{Reciprocal Tandem}\label{sec:reciprocal tandem}

In this section, we introduce the concept of \emph{reciprocal tandem}. Before attending to the rigorous definition and the use cases of reciprocal tandem, we start with a toy example that captures the main idea behind such a definition.   

\subsubsection{Example 1} Consider a MU that is transmitting a pilot signal using only the first digital RF chain and a single beamforming vector $\b$ in the uplink and an AP which is receiving the pilot with beamforming vector $\f$ on the first digital RF chain. Therefore, the received signal at the AP is given by
\begin{align}
    &\y_{\text{ap}} = \r_{11}\f^T\R_2\H^T\T'_2\b \t'_{11} s + \Z
\end{align}
where $s$ is the transmitted pilot signal. Hence, the AP can estimate the end-to-end uplink channel as follows, 
\begin{align}
     \r_{11}\f^T\R_2\H^T\T'_2\b\t'_{11} \doteq \Tilde{\y}_{\text{ap}}.
\end{align}
%where $R_1$ and $R_2$ are the digital and analog receiver calibration matrices at the BS, $T'_1$ and $T'2$ are the digital and analog transmitter calibration matrices at the MU, respectively.
Multiplying by the inverse of the digital chain calibration coefficients we have
\begin{align}
     \f^T\R_2\H^T\T'_2\b \doteq\r_{11}^{-1}\Tilde{\Y}_{\text{ap}}\t'^{-1}_{11}.
\end{align}
Transposing both sides will give, 
\begin{align}
     \b^T\T'_2\H\R_2\f \doteq\t_{11}'^{-1}\Tilde{\Y}^T_{\text{ap}}\r^{-1}_{11}
\end{align}
and multiplying by $\r'_{11}$ from left and $\t_{11}$ from right we have
\begin{align}
     \r'_{11} \b^T\T'_2\H\R_2\f \t_{11} \doteq \r'_{11} \t_{11}'^{-1}\Tilde{\Y}^T_{\text{ap}} \r^{-1}_{11} \t_{11}
\end{align}
where the last equation can be rephrased as follows, 
\begin{align}
    \r'_{11} \b^T\left(\R_2'^{-1}\T_2'\right)\R'_2\H\T_2\left(\R_2\T_2^{-1}\right) \f \t_{11} = \r'_{11} \t_{11}'^{-1}\Tilde{\Y}_{\text{ap}}^T\r^{-1}_{11} \t_{11} \label{eq:ex1}
\end{align}
where we utilized the fact that the analog calibration matrices are diagonal and therefore the matrix multiplication for them is commutative. Comparing \eqref{eq:ex1} with the received signal at the MU given by
\begin{align}
    &\Y_{\text{mu}} = \r'_{11}\check{\b}^T\R_2'\H\T_2\check{\f}\t_{11} s + \Z 
\end{align}
one can interpret that the downlink channel has been indeed estimated by the AP through this process for the case that the analog transmit beamformer $\check{\b} = \left(\R_2'\T_2'^{-1}\right)^{-1} \b $ is used by the AP and the analog receive beamformer $\check{\f} = \left(\R_2\T_2^{-1}\right) \f$ is used by the MU. This example reveals the fact that to use the beamformers $\check{\b}$ and $\check{\f}$ by the AP and MU in downlink transmission, one needs to estimate the uplink channel by using the reciprocal tandems of $\check{\b}$ and $\check{\f}$ given by $\b = (\R'_2\T_2'^{-1}) \check{\b}$ and $\f = (\R_2\T_2^{-1})^{-1} \check{\f}$, respectively.

\begin{defn}
The \emph{reciprocal tandem} of a receive beamformer $\f$ at a communication node with the receive analog RF chain calibration matrix $\R_2$ and the transmit analog RF calibration matrix $\T_2$ is given by $\check{\f} = \R_2 \T^{-1}_2 \f$. 
\end{defn}
Alternatively, one may define the reciprocal tandem as a pair of transmit and receive beamformers used in transmission and reception by the same node which allows the reciprocity to hold in spite of the existing mismatch between the transmit and receive analog RF chains. Obviously, using either definition, the tandem pair is defined per node and only depends on the analog calibration matrices and its definition does not depend on the channel or other possible mismatch or calibration parameters at the other nodes. Two notes are in order here. First, based on the definition of the reciprocal tandem of a receive beamformer, a reciprocal tandem of a transmit beamformer $\f$ is given by $\check{\f} = (\R_2 \T^{-1}_2)^{-1} \f$. Second, if $\F$ is a matrix, its reciprocal tandem $\check{\F}$ is a same-size matrix where each column is the reciprocal tandem of the corresponding column of $\F$.

%% file: downlink.tex
\section{Reciprocity-based DL Channel Estimation }
\label{channel estimation}

We note that the ultimate goal of reciprocity calibration is indeed finding the downlink channel estimate based on the uplink channel estimation using the pilots transmitted by the users. In this section, we first propose a two-step approach for calibration between any two nodes and then proceed to downlink channel estimation. 

\subsection{Reciprocity Calibration between Two Nodes}\label{subsec:lemmas}

\begin{algorithm}[t]
\caption{\textit{Digital Chain Calibration}}
\label{alg:dig}
 \begin{algorithmic}[1]
 \renewcommand{\algorithmicrequire}{\textbf{Input:}}
 \renewcommand{\algorithmicensure}{\textbf{Output:}}
 \REQUIRE $s, \F = \f_1, \B = [\b_1, \b_1, \ldots, \b_1].$
 \STATE \textbf{For} $i = 1 \text{ to } N$:
 \STATE \quad Tx: Send $s$ with digital chain $t_{1n}$ and beamformer $\F$.
  \STATE \quad Rx: Receive vector $\y'_i$ on all RF chains with beamformer\\ \quad $\B$ according to: 
  \begin{align}
    \y'_i = \R_1'\B^T\R_2'\H\T_2\f_1 t_{1i} s + z_{DL} \nonumber
\end{align}
 \STATE \textbf{End For}
 \STATE Define $h \doteq \b_1^T \R_2'\H\T_2\f_1$. \textbf{Solve} 
 \begin{align}
    \{t_{1i}, r'_{1i}\}_{i=1}^N = \arg \min_{\{t_{1i}, r'_{1i}\}_{i=1}^N} \sum_{i=1}^N\sum_{j=1}^N\left\|y'_{ij} - r'_{1i} h t_{1j} \right\|^2\nonumber
\end{align}\label{step_solve}

\STATE \textbf{Return} $\T_1 = \{t_{1i}\}_{i=1}^N$, and $\R'_1 = \{r'_{1i}\}_{i=1}^N$.
\end{algorithmic}
\end{algorithm}

\begin{algorithm}[t]
\caption{\textit{Analog Chain Calibration}}
\label{alg:analog}
 \begin{algorithmic}[1]
 \renewcommand{\algorithmicrequire}{\textbf{Input:}}
 \renewcommand{\algorithmicensure}{\textbf{Output:}}
 \REQUIRE $s, \R_1, \T_1, \R'_1, \T'_1$, Full-Rank $\F$, Full-Rank $\B$ such that \\ \quad $\F = \left[\f_1, \f_2, \ldots, \f_M\right], \B = [\b_1, \b_2, \ldots, \b_M]$
 \STATE \textbf{For} $\f_i, i = 1 \text{ to } M$:\label{for_start}
  \STATE \quad \textbf{For} $k=0\ldots\ceil{M/N}-1:$
  \STATE \quad \quad Tx: Send $s$ on digital chain $t_{11}$ with beamformer $\f_i$
  \STATE \quad \quad Rx: Receive $\y'_{ki}$ on all digital chains with beamformer \\\quad\quad$\B_k = \left[\b_{1+k N}, \b_{2+k N}, \ldots, \b_{N+kN}\right]$ according to:
  \begin{align}
    & \y'_{ki} = \R_1'\B_k^T\R_2'\H\T_2\f_i t_{11}s + \z'_{ki}\nonumber
\end{align}
  \STATE \quad \textbf{End For}
 \STATE \textbf{End For}\label{for_end}
 \STATE Stack the rows of $\R'_1\B_k,$ $k=1\ldots \ceil{M/N}-1$ to get $r'_{11}\tilde{\B}$
 \STATE Obtain $\Y' = r'_{11} \Tilde{\B}^T\R_2'\H\T_2\F t_{11}s' + \Z'$.
\STATE Repeat steps~\ref{for_start}-\ref{for_end} for the uplink direction and obtain
\begin{align}
    \Y = r_{11} \Tilde{\F}^T\R_2\H^T\T'_2\B t'_{11}s + \Z. \nonumber
\end{align}\label{calib_done}
 \STATE Let $\beta = {r_{11}^{-1} t_{11}'^{-1}} / ({ r_{11}'^{-1} t_{11}'^{-1}})$, $\alpha_i = r_{2i} t_{2i}^{-1}$, $\alpha_i' = r'_{2i} t_{2i}'^{-1}$.
 \STATE Obtain matrices $\X \doteq \Tilde{\B}^{-T}\Tilde{\Y}'\F^{-1}$ and $\Z \doteq \B^{-T}\Tilde{\Y}^T\Tilde{\F}^{-1}$.
\STATE \textbf{Solve}
\begin{align}
    \{\alpha_i, \alpha'_i\}_{i=1}^M = \arg \min_{\alpha_i ,\alpha'_j, i,j\in[M]} \sum_{i=1}^M\sum_{j=1}^M\left\|x_{ij}\alpha_j -\beta z_{ij}\alpha'_i\right\|^2\nonumber
\end{align}\label{analog_formula}
\STATE \textbf{Return} $\R_2\T_2^{-1} = diag\{\alpha_i\}_{i=1}^M$, $\R_2' \T_2'^{-1} = diag\{\alpha'_i\}_{i=1}^M$  
\end{algorithmic}
\end{algorithm}

Consider the calibration between two nodes $S$ and $S'$ from the distributed MIMO system that are capable of hybrid beamforming. Without loss of generality assume each node employs $M$ analog and $N$ digital RF chains, respectively. Let  $\T_1$, $\R_1, \T'_1$, and $\R_1'$ be the digital calibration matrices of the nodes $S$ and $S'$, respectively. Similarly, let $\R_2 \T_2^{-1}$ and $\R_2' \T_2'^{-1}$ be the calibration matrices corresponding to the analog chains of the nodes $S$ and $S'$, respectively. The calibration process occurs  in two steps. Algorithm~\ref{alg:dig} shows how the calibration for the digital chain is performed. Once the $N$ rounds of communications between the two nodes is complete and $\y_i, i=1,\ldots,N$ is obtained, the least-square optimization problem in step~\ref{step_solve} is solved. Note that since $h$ is an unknown parameter, the digital reciprocity matrices $\T_1$ and $\R'_1$ are determined up to an unknown scaling factor. Repeating the above process, the digital reciprocity matrices $\T'_1$ and $\R_1$ will be computed up to an unknown scaling factor. Algorithm~\ref{alg:analog} presents the reciprocity calibration procedure for the analog chain. Forming the invertible matrices $\F$ and $\B$ and following steps~\ref{for_start}-\ref{calib_done} we obtain matrices $\Y$ and $\Y'$ for the uplink and downlink direction. Accordingly, we can compute the channel $\H$ using these observations in two ways. 
\begin{align}
    \Tilde{\H}_1 = (r'_{11} t_{11})^{-1}\R'^{-1}_2\Tilde{\B}^{-T}\Tilde{\Y}'\F^{-1}\T_2^{-1}\label{eq:ch1} \\
    \Tilde{\H}_2 = (r_{11} t'_{11})^{-1}\T'^{-1}_2\B^{-T}\Tilde{\Y}^T\Tilde{\F}^{-1}\R_2^{-1}\label{eq:ch2}
\end{align}

By setting equal the last two matrices element-wise, we formulate the analog chain reciprocity calibration problem as a least-square optimization problem as in step~\ref{analog_formula} and obtain the analog calibration matrices $\R_2\T_2^{-1}$ and $\R'_2\T_2'^{-1}$ up to an unknown scaling factor. The estimated reciprocity matrices are used in DL channel estimation. We first consider DL channel estimation for a single AP and then extend our analysis to the case of multiple APs.  

\subsection{Downlink Channel Estimation with a Single AP}\label{subsec:single}
%To this end, we consider two examples. Before zooming on the examples note that when the transmitter is using the transmit beamforming matrix $\F$ and the receiver is using the receive beamforming matrix $\B$, 
Using $\F_{\text{mu}}$ as the transmit beamformer at the AP and $\B_{\text{mu}}$ as the receive beamformer at the MU in the DL direction, and similarly using $\B_{\text{bs}}$ as the transmit beamformer at the MU and $\F_{\text{bs}}$ as the receive beamformer at the AP in the UL direction the signal model at the MU and the AP is given by, 
\begin{align}
    &\Y_{\text{mu}} = \R'_1\B_{\text{mu}}^T\R_2'\H\T_2\F_{\text{mu}}\T_1s + \Z_{mu} \label{eq:dl_eq}\\
    &\Y_{\text{ap}} = \R_1\F_{\text{ap}}^T\R_2\H^T\T'_2\B_{\text{ap}}\T'_1s + \Z_{ap}
\end{align}

Since the digital calibration matrices are known up to a scaling factor, the problem reduces to finding the effective DL channel $\H^{\text{eff}}_{\text{DL}} \doteq \R_2'\H\T_2$ based on observations of $\Y_{ap}$ which involves the effective uplink channel $\H^{\text{eff}}_{\text{UL}} \doteq \R_2\H^T\T'_2$. By simple algebra, one can find the effective DL channel matrix in terms of the effective UL channel matrix as
\begin{align}
    \H^{\text{eff}}_{\text{DL}} = \left(\R'_2\T_2'^{-1}\right)  (\H^{\text{eff}}_{\text{UL}})^T \left(\R_2\T_2^{-1}\right)^{-1}\label{dl_eff}
\end{align}

Now, it remains to estimate $\H^{\text{eff}}_{\text{UL}}$. We select $M_{\text{mu}}$ transmit beamformer $\b_i$, $i = 1, \ldots, M_{\text{mu}}$ and $M_{\text{ap}}$ receive beamformers $\f_i, i = 1, \ldots, M_{\text{ap}}$ from the transmit and receive codebook, respectively, such that the matrices $\F = \left[\f_1, \f_2, \ldots, \f_{M_{\text{ap}}}\right]$ and $\B = \left[\b_1, \b_2, \ldots, \b_{M_{\text{mu}}} \right]$ are full rank. For each $\b_i$, $i = 1, \ldots, M_{\text{mu}}$, we perform $k = 0, 1, \ldots, \ceil{M_{\text{ap}}/N_{\text{ap}}}-1$ transmissions with transmit beamformers $\b_i$ from chain 1, and receive beamformers $\f_{1+k \ceil{M_{\text{ap}}/N_{\text{ap}}}}, \f_{2+k \ceil{M_{\text{ap}}/N_{\text{ap}}}}, \ldots, \f_{N_{\text{ap}}+k \ceil{M_{\text{ap}}/N_{\text{ap}}}}$ for the $k^{th}$ transmissions with the beamformer $\b_i$, where $\b_m = \b_1$ for $m > M_{\text{mu}}$. Hence, after $M_{\text{mu}} \ceil{M_{\text{ap}}/N_{\text{ap}}}$ transmission we gather $M_{\text{mu}}M_{\text{ap}}$ observations that we arrange in a $M_{\text{mu}} \times M_{\text{ap}}$ matrix $\Y$.  Using the uplink model \eqref{UL_tran}, we can write
\begin{align}
    & \y_{ki} = \R_1\F_k^T\R_2\H\T'_2\b_i t'_{11}s + \z'_{ki}, 
\end{align}
for all $k=1, \ldots, \ceil{M_{\text{ap}}/N_{\text{ap}}}$ and $i = 1, \ldots, M_\text{mu}$ where the estimated values for $\R_1$ is already computed up to a scaling factor $r_{11}$. The $j^{th}$, $j = 1, \ldots, N_{\text{ap}}$ row of the matrix product $\R_1\F_k$ for $k = 1, \ldots, \ceil{M_{\text{ap}}/N_{\text{ap}}}$ is given by a row vector 
\begin{align}
    r_{1j} \f_{N_{\text{ap}}(k-1)+j}^T = r_{11} (r_{1j}/r_{11}) \f_{N_{\text{ap}}(k-1)+j}^T \doteq r_{11} \tilde{\f}_m^T
\end{align}
where $m = N_{\text{ap}}(k-1)+j$. Stacking these row vectors $\tilde{\f}_m^T$ for $m = 1, \ldots, M_{\text{ap}}$ results in an $M_{\text{ap}} \times M_{\text{ap}} $ matrix $\tilde{\F}$. Considering $\F = [\f_1, \f_2, \ldots, \f_{M_{\text{ap}}}]$ which consists of all $M_{\text{ap}}$ transmit beamforming vectors and the modified received beamforming vectors in $\tilde{\F}$, we would get an aggregate $M_{\text{ap}} \times M_{\text{ap}}$ observation matrix $\Y$ as
\begin{align}
    \Y = r_{11} \Tilde{\F}^T\R_2\H^T\T'_2\B t'_{11}s + \Z. \label{secon-level_dl}
\end{align}
% Following the same approach as in Section~\ref{sec:analog chain}, in the UL direction we require $M_{\text{mu}}\ceil{M_{\text{ap}}/N_\text{ap}}$ pilot transmissions which generates $M_{\text{ap}}M_{\text{mu}}$ observations. 
% %
%Following the same sequel, denote the stacked beamforming matrices that are extracted from the transmit and receive codebook by $\check{\F}$ and $\check{\B}$ and the matrix of noisy observations by $\Tilde{\Y}$. The $\check{\F}$ and $\check{\B}$ matrices are full-rank and therefore invertible. Similar to \eqref{secon-level_dl}, it holds in the UL direction that, 
%\begin{align}
%    \Tilde{\Y} = \R_1\Tilde{\F}^T\R_2\H^T\T'_2\Tilde{\B}\T'_1
%\end{align}
%
% Following the same sequel, the equations \eqref{secon-level_dl} can be written as
% \begin{align}
%     \Y = r_{11} \Tilde{\F}^T\R_2\H^T\T'_2\B t'_{11}s + \Z. \label{secon-level_dl_new}
% \end{align}
where $\B$ is comprised of putting $M_{\text{mu}}$ transmit beamforming vectors used by MU each in separate transmissions in one row to get an $M_{\text{mu}} \times M_{\text{mu}}$ matrix and $\Tilde{\F}^T$ is comprised of the transpose of the modified received beamforming vectors in one column to get an $M_{\text{ap}} \times M_{\text{ap}}$ matrix. We note that the $M_{\text{mu}}$ transmit beamforming vectors, and the $M_{\text{ap}}$ receive beamforming vectors are chosen such that $\B$ and $\F$ (and hence, $\tilde{\F}$) are invertible. Let $\Tilde{\Y}$ be the estimation of $r_{11} \Tilde{\F}^T\R_2\H^T\T'_2\B t'_{11}$. Hence, the effective uplink channel matrix can be written as
\begin{align} 
    \H^{\text{eff}}_{\text{UL}} =\left( \left(\r_{11}\Tilde{\F}^T\right)^{-1}\Tilde{\Y}\left({\B}\t'_{11}\right)^{-1}\right)^T\label{ul_eff}
\end{align}

\subsection{Downlink Channel Estimation with Multiple APs}
Contrary to the case of beamforming from a single AP, where achieving the calibration between the MU and the AP was enough to obtain the downlink channel estimate by combining the equations \eqref{dl_eff} and \eqref{ul_eff}, we show that when multiple APs co-operate towards jointly serving an MU a third calibration step must be performed between the co-operative APs in addition to the two calibration steps set forth in section~\ref{subsec:lemmas}. Let us consider the scenario involving two collaborating APs, namely, AP$_1$, and AP$_2$. The extension of the results to the case of multiple APs is straightforward. 
% The APs will first perform calibration among themselves using the proposed two-step calibration approach in section~\ref{sec:problem}, and obtain their digital and analog calibration matrices up to an unknown scaling factor. 
Suppose each of the APs estimates its DL channel to the MU separately, by performing calibration between the AP and the MU as discussed in section~\ref{subsec:single}. Since the calibration matrices are only known up to a scaling factor, the DL received signal model between each AP and the MU is given by, 
\begin{align}
    &\y_{\text{DL, 1}} = c_1\H_{\text{DL,1}}\s + \z_{DL,1}\\
    &\y_{\text{DL, 2}} = c_2\H_{\text{DL,2}}\s + \z_{DL,2}
\end{align}
where $c_1$ and $c_2$ are the unknown coefficients of the estimated downlink channel between AP$_1$, AP$_2$ and the user, respectively. The overall downlink channel from both APs is obtained by collecting all the column vectors of $c_1\H_{\text{DL,1}}$ and $c_2\H_{\text{DL,2}}$ into a single channel matrix  $\H_{\text{DL}} = [c_1\H_{\text{DL,1}}, c_2\H_{\text{DL,2}}]$. We note that in general the coefficients $c_1$ and $c_2$ are unknown and uncorrelated parameters. Therefore, in order to perform co-operative beamforming we need to at least estimate the ratio $c_2/c_1$. Following equation~\eqref{eq:dl_eq}, it can be easily inferred that 
\begin{align}
    &c_1 = r^{mu}_{11}t^{1}_{11}\sigma^{mu}_2(\sigma^{1}_2 t^{mu}_{11} r^{1}_{11})^{-1}\\
    &c_2 = r^{mu}_{11}t^{2}_{11}\sigma^{mu}_2(\sigma^{2}_2 t^{mu}_{11} r^{2}_{11})^{-1}    
\end{align}
where $r^{mu}_{11}$, $t^{mu}_{11}$, and $\sigma^{mu}_2$ are the unknown scaling factors (embedded in the first elements) of the calibration matrices $\R^{mu}_1, \T^{mu}_1$, and $\R^{mu}_2 (\T^{mu}_2)^{-1}$ for the MU. Similarly, $r^{k}_{11}$, $t^{k}_{11}$, and $\sigma^{k}_2$ are the unknown scaling factors (embedded in the first elements) of the calibration matrices $\R^{k}_1, \T^{k}_1$, and $\R^{k}_2 (\T^{k}_2)^{-1}$ for the $k$-th AP, $k = 1, 2$. Define $c\doteq c_1/c_2$, we have 
\begin{align}
    c &= \frac{r^{mu}_{11}t^{1}_{11}\sigma^{mu}_2(\sigma^{1}_2 t^{mu}_{11} r^{1}_{11})^{-1}}
    {r^{mu}_{11}t^{2}_{11}\sigma^{mu}_2(\sigma^{2}_2 t^{mu}_{11} r^{2}_{11})^{-1}} = \frac{t^{1}_{11}(\sigma^{1}_2 r^{1}_{11})^{-1}} {t^{2}_{11}(\sigma^{2}_2 r^{2}_{11})^{-1}} \label{c_coeff}
\end{align}

% \begin{align}
%     &c_1 = r^{(mu)}_{11}t^{(1)}_{11}\sigma^{(mu)}_2(\sigma^{(1)}_2)^{-1}(t^{(mu)}_{11})^{-1}(r^{(1)}_{11})^{-1}\\
%     &c_2 = r^{(mu)}_{11}t^{(2)}_{11}\sigma^{(mu)}_2(\sigma^{(2)}_2)^{-1}(t^{(mu)}_{11})^{-1}(r^{(2)}_{11})^{-1}    
% \end{align}
% where $r^{(mu)}_{11}$, $t^{(mu)}_{11}$, and $\sigma^{(mu)}_2$ are the unknown scaling factors (embedded in the first elements) of the calibration matrices $\R^{mu}_1, \T^{mu}_1$, and $\R^{mu}_2 (\T^{mu}_2)^{-1}$ for the MU. Similarly, $r^{(k)}_{11}$, $t^{(k)}_{11}$, and $\sigma^{(k)}_2$ are the unknown scaling factors (embedded in the first elements) of the calibration matrices $\R^{(k)}_1, \T^{(k)}_1$, and $\R^{(k)}_2 (\T^{(k)}_2)^{-1}$ for the $k$-th AP, $k = 1, 2$. Define $c\doteq c_1/c_2$, we have 
% \begin{align}
%     c &= \frac{r^{(mu)}_{11}t^{(1)}_{11}\sigma^{(mu)}_2(\sigma^{(1)}_2)^{-1}(t^{(mu)}_{11})^{-1}(r^{(1)}_{11})^{-1}}{r^{(mu)}_{11}t^{(2)}_{11}\sigma^{(mu)}_2(\sigma^{(2)}_2)^{-1}(t^{(mu)}_{11})^{-1}(r^{(2)}_{11})^{-1}} \nonumber \\
%     & = \frac{t^{(1)}_{11}(\sigma^{(1)}_2)^{-1}(r^{(1)}_{11})^{-1}}{t^{(2)}_{11}(\sigma^{(2)}_2)^{-1}(r^{(2)}_{11})^{-1}}\label{c_coeff}
% \end{align}

It is observed that the coefficient $c$ is not a function of the calibration parameters of the MU and only depends on the calibration parameters of the APs. Suppose the two-step calibration process between the APs is performed and the two APs have estimated the calibration matrices up to a scaling factor. We show that the coefficient $c$ in \eqref{c_coeff} can be obtained by taking a third step in the calibration process. In the following, we show employing the notion of reciprocal tandem may facilitate estimation of this ratio efficiently by only two pilot transmissions, one from each AP. 
% In order to find $c$ we have to calculate the last ratio within the calibration process between the two APs. To this end, we use the reciprocal tandem property as follows.
AP$_1$ transmits a pilot signal from its first digital RF chain using the beamforming vector $\f^{1}_1$ that is received by AP$_2$ on its first digital RF chain using beamforming vector $\b^{2}_1$. In the reverse direction, AP$_2$ transmits a pilot signal from its first digital RF chain using $\b^2_2$ which is the scaled version of the reciprocal tandem of $\b^2_1$ and AP$_1$ receives the signal in its first digital RF chain using the beamforming vector $\f^1_2$ which is the scaled version of the reciprocal tandem of $\f^1_1$. Since $\R^2_2(\T^2_2)^{-1}$ is known up to an scaling factor, the scaling factor is taken as its first diagonal element which is denoted by $\sigma^2_2$ and hence the estimated matrix can be written as $\R^2_2(\T^2_2)^{-1} / \sigma^2_2$. As a result, we can compute $\b^2_2 = (\sigma^2_2)^{-1} \check{\b}^2_1 = (\sigma^2_2)^{-1} \R^2_2(\T^2_2)^{-1}\b^2_1$. Similarly, we can compute $\f^1_2 = (\sigma^1_2) \check{\f}^1_1 = \left((\sigma^1_2)^{-1} \R^1_2 (\T^1_2)^{-1} \right)^{-1} \f^1_1$. Let the observations in both directions be $\Tilde{y}_{12}$ and $\Tilde{y}_{12}$. We get,  
% Let us denote end-to-end channel the first digital RF chains from AP$_k$ to AP$_l$ by $\tilde{y}_{kl}$. We have
%are used in the calibration process between the two APs in the transmission direction from AP$_1$ to AP$_2$, i.e AP$_1$ uses the transmit beamforming vector $\f^{(1)}_1$ on the first digital chain with mismatch coefficient $t^{(1)}_{11}$ and AP$_2$ uses the receive beamforming vector $\b^{(2)}_1$ on the first digital chain with mismatch coefficient $r^{(2)}_{11}$. In the reverse direction we use the reciprocal tandems of each of the given beamforming vectors, i.e. AP$_2$ uses $\b^{(2)}_2 = \check{\b^{(2)}_1} = \left(\sigma^{(2)}_1\R^{(2)}_2(\T^{(2)}_2)^{-1}\right)\b^{(2)}_1$ on the first digital chain with the mismatch coefficient $t^{(2)}_{11}$  and AP$_1$ uses $\f^{(1)}_2 = \check{\f^{(1)}_1} = \left(\sigma^{(1)}_1\R^{(1)}_2(\T^{(1)}_2)^{-1}\right)^{-1}\f^{(1)}_1$ on the first RF chain $r^{(1)}_{11}$. We have 
\begin{align}
    &\Tilde{y}_{12} = r^2_{11} \left(\b_1^2\right)^T \R^2_2 \H \T^1_2 \f_1^1 t^1_{11}\\
    &\Tilde{y}_{21} = r^1_{11} \left(\f^1_2\right)^T \R^1_2 \H^T \T^2_2\b^2_2 t^2_{11}\label{eq:recip}
\end{align}

Transposing the RHS of \eqref{eq:recip} and replacing $\check{\b^2_2}$ and $\check{\f^{1}_2}$ with their equivalent values, we can rewrite it as
\begin{align}
    \Tilde{y}_{21} = t^2_{11} (\sigma^2_2)^{-1} \left(\b^2_{1}\right)^T\R^2_2\H\T^1_2 \sigma^1_2 \f^1_{1}r^1_{11}
\end{align}

The ratio of the scalar observations $\Tilde{y}_{12}$ and $\Tilde{y}_{21}$, would give 
% \begin{align}
%     \Tilde{y}_{12}/\Tilde{y}_{21} = \left(r_{11}'t_{11}'^{-1}\right)\left(r_{11}t_{11}^{-1}\right)^{-1}\left(r_{21}'t_{21}'^{-1}\right)\left(r_{21}t_{21}^{-1}\right)^{-1}
% \end{align}
\begin{align}
    \Tilde{y}_{12}/\Tilde{y}_{21} = \frac{t^1_{11}(\sigma^1_2)^{-1}(r^1_{11})^{-1}}{t^2_{11}(\sigma^2_2)^{-1}(r^2_{11})^{-1}}
\end{align}
that can directly be used to estimate the parameter $c$. It is easy to extend this analysis to the case of multiple APs by picking one AP as the reference and calibrating the rest of APs with respect to the reference one.

%% file: evaluation.tex
\section{Performance Evaluation}
\label{sec:evaluation}

\begin{figure*}
\centering
\begin{minipage}{0.32\textwidth}
\includegraphics[width=\textwidth]{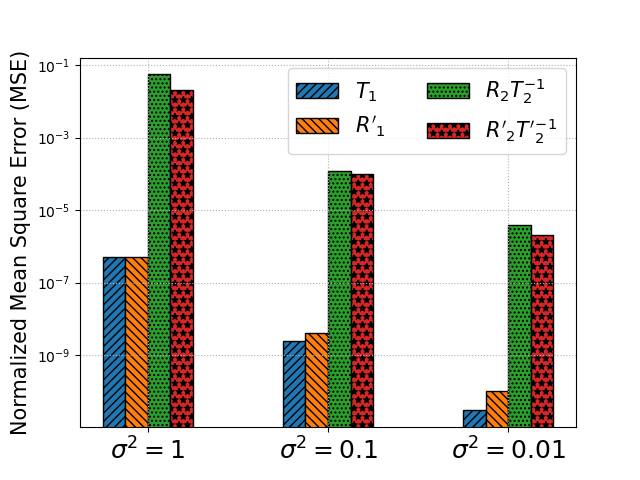}
\caption{Cal. MSE vs. Noise Variance }    
\label{fig:MSE}
\end{minipage}
\begin{minipage}{0.33\textwidth}
    \includegraphics[width=\textwidth]{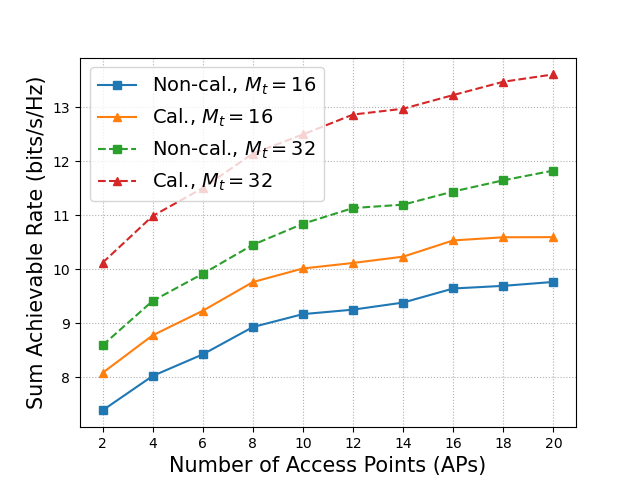}
    \caption{Sum  Rate under varying $K$}
    \label{fig:ZFBF_AP}
\end{minipage}
\begin{minipage}{0.33\textwidth}
    \includegraphics[width=\textwidth]{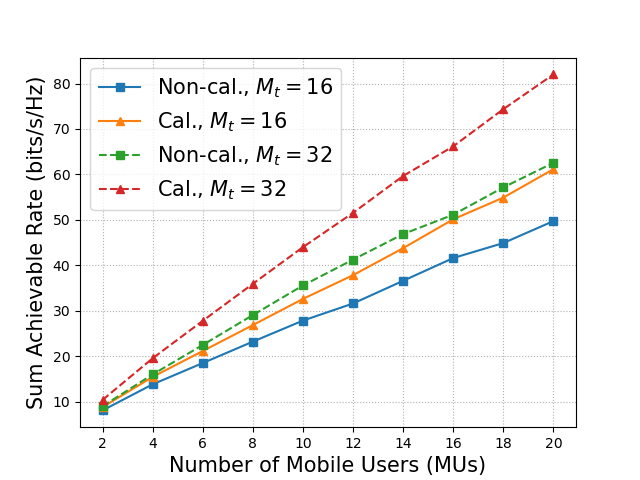}
    \caption{Sum  Rate under varying $U$}
    \label{fig:second}
\end{minipage}
% \caption{Sum Rate Performance}
\end{figure*}

\begin{figure*}
\centering
\begin{subfigure}{0.32\textwidth}
    \includegraphics[width=\textwidth]{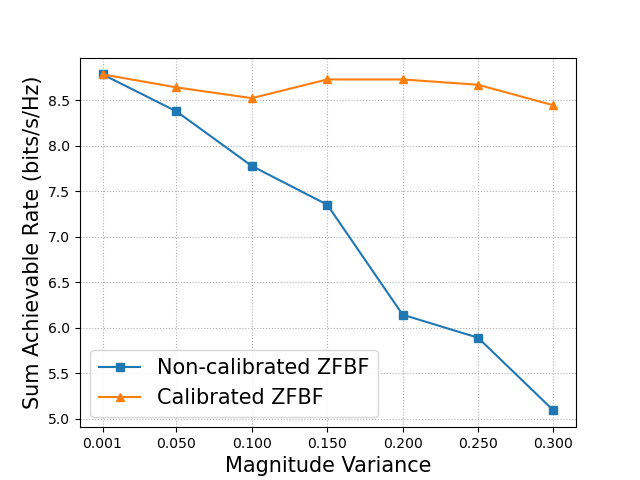}
    \caption{Varying mismatch magnitude}
    \label{fig:first}
\end{subfigure}
\hfill
\begin{subfigure}{0.32\textwidth}
    \includegraphics[width=\textwidth]{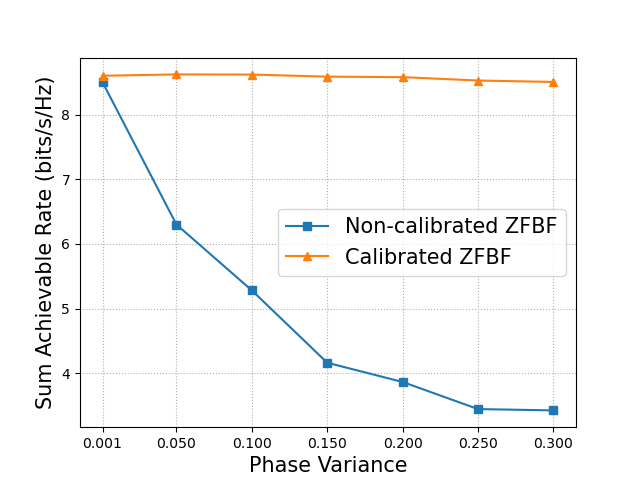}
    \caption{Varying mismatch phase }
    \label{fig:second}
\end{subfigure}
\hfill
\begin{subfigure}{0.32\textwidth}
    \includegraphics[width=\textwidth]{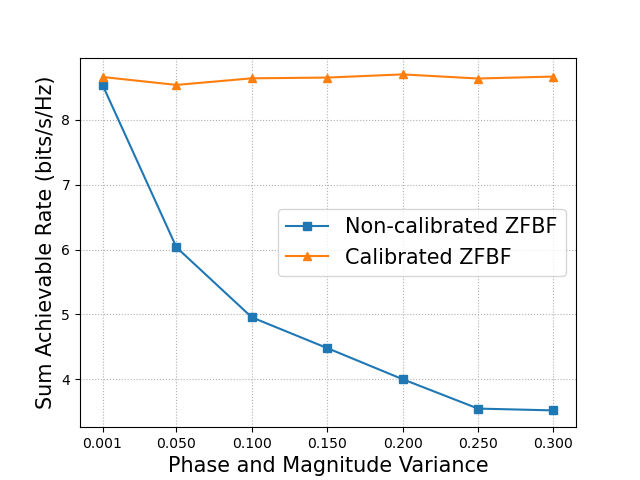}
    \caption{Varying mismatch phase and magnitude }
    \label{fig:third}
\end{subfigure}
        
\caption{Sum Rate Performance under varying Mismatch with $K=U=2$ }
\label{fig:figures}
\end{figure*}

% In this section, we verify the effectiveness of our channel reciprocity calibration procedure in boosting the performance of CHBF, by simulating a ZFBF scenario in a distributed MIMO system. In what follows, we first describe the simulation setup and then proceed to the analysis of the numerical results. 

\subsection{Simulation Setup and Parameters}

Multiple APs capable of hybrid beamforming jointly serving a group of single-antenna MUs are considered. Throughout the experiments, the APs may employ $M_t = 32, 64$ antennas and $N_t = M_t/4$ digital RF chains. We consider a multi-path channel model with $L=4$ paths. We assume the gain of each path $\alpha_\ell, \ell= 1,\ldots,4$ follows a Gaussian distribution with mean zero and variance $\sigma^2_\alpha =1$. The AoAs/AoDs are sampled from a uniform distribution $\left\{\theta_k, \phi_k\right\} \sim \mathcal{U}(-\pi / 2, \pi / 2)$. Further, we assume the reciprocity mismatch gains follow a log-normal distribution, i.e. $\left\{\ln \left|t_{i, n}\right|, \ln \left|r_{i, n}\right|, \ln \left|t'_{i, n}\right|, \ln \left|r'_{i, n}\right|\right\} \sim \mathcal{N}(0,\sigma), i\in \{1,2\}, n = 1\ldots N$, for standard deviation $\sigma$. The phase of the mismatch parameters follows a uniform distribution $\left\{\angle t_{i, n}, \angle r_{i, n}, \angle t'_{i, n}, \angle r'_{i, n}\right\} \sim \mathcal{U}(-\sigma, \sigma)$. Our tests run on an Intel i9 CPU at 2.3 GHz and 16 GB of main memory.

\subsection{Co-operative Zero-forcing Beamforming (ZFBF)}
Consider a multi-AP system with $K$ APs each employing $N$ antennas that is serving a group of $U$ MUs under ZFBF. The received signal at user $u$ is given by, 
\begin{align}
    \mathbf{y}={\H^T_{\text{DL}}} \mathbf{W D s}+\mathbf{z}_{\text{DL}}
\end{align}
where $\H = \left[\h_1, \ldots, \h_{U}\right] \in \mathbb{C}^{KN\times U}$ represents the downlink channel matrix, $\W = \left[\w_1, \ldots, \w_{U}\right] \in \mathbb{C}^{KN\times U}$ denotes the precoding matrix, $\mathbf{D}=\operatorname{diag}\left(\sqrt{P_1}, \ldots, \sqrt{P_U}\right) \in \mathbb{R}^{U\times U}$ is the diagonal power matrix, and $\s = \left[s_1, \ldots, s_{U}\right]^T \in \mathbb{C}^{U}$ is the transmit signal. Under zero-force precoding, for every user $u$, the corresponding precoder $\w_u$ is orthogonal to all the channel vectors $\h_v$ associated with other users $v \neq u$. This way the interference resulting from the desired transmission for a user is suppressed for other users when the aggregate signal passes through the channel. In the matrix form, the orthogonality condition can be stated as $\H^T\W = \Q$ where $\Q$ is usually picked as the Identity matrix. Therefore, the optimal $\W$ can be found as the pseudo-inverse of matrix $\H^T$ as $\W = \mathbf{H}\left(\mathbf{H}^T \mathbf{H}\right)^{-1}$. In this experiment, we simulate a cooperative ZFBF scenario in a multi-user setup with and without calibration. We consider the achievable sum rate of the users to compare the performance of ZFBF with and without calibration. The sum rate is computed as $r = \sum_{u=1}^U \log(1 + SNR_u)$, where $SNR_u$ is the output signal-to-noise ratio corresponding to user $u$. When no calibration technique is in place, the beamforming precoders are decided based on the uplink channel estimate which is not necessarily a reasonable approximation of the downlink channel depending on the level of imperfections. This will result in lower output SNR and therefore lower user sum rate compared to the case where calibration is in place and the downlink channel is accurately estimated.  Fig.~\ref{fig:MSE} shows the high accuracy of our calibration technique under different noise levels by depicting the mean squared error (MSE) of the estimated calibration matrices where all the estimated matrices are normalized with respect to their first elements. Fig.~\ref{fig:ZFBF_AP} shows how the user sum rate evolves by increasing the number of APs when there are 2 MUs targeted. seen that the channel reciprocity calibration enhances the user sum rate by $20 \%$ and $30\%$ for $M_t = 16$ and $M_t = 32$, respectively, when averaged over different cases on the number of APs in the distributed massive MIMO network. Fig.~\ref{fig:second} investigates the same effect as the number of MUs increases where there are 2 APs in place. We observe that as the number of targeted MUs increases the impact of calibration becomes more visible. We conclude that channel reciprocity calibration in multi-user scenarios is drastically important, and its importance gains more attention in densely populated environments.  Fig.~\ref{fig:figures} measures the impact of our scheme on the performance of ZFBF at different levels of reciprocity mismatch where there are $2$ MUs and $2$ APs. In figures~\ref{fig:first} and \ref{fig:second}, the magnitude and the phase of reciprocity mismatch are changing such that the intensity of the hardware imperfection increases. We observe that as these imperfections worsen (both in magnitude and phase), the performance of the non-calibrated ZFBF degrades and the users achieve a lower rate. however, when our technique is in place the achieved rate remains constant. Fig.~\ref{fig:third} shows this effect when the magnitude and the phase of the mismatch coefficients tend to deviate from their ideal values simultaneously. The trend in this figure is closer to the trend in Fig.~\ref{fig:second} which indicates that the CHBF is more vulnerable to phase mismatch.

%% file: conclusion.tex
\section{Conclusions}
We proposed a channel reciprocity calibration technique for collaborative hybrid beamforming in a distributed MIMO system. We presented the novel concept of reciprocal tandem and utilized that in calibrating the cluster of APs and estimating the downlink channel. 
\label{sec:conclusions}